# Fast Optimization of Impulsive Perturbed Orbit Rendezvous with Finite Iterations

An-yi Huang*

*Xi'an Satellite Control Center, 710043 Xi'an, People's Republic of China*

*National University of Defense Technology, 410073 Changsha, People's Republic of China*

Ya-zhong Luo†

*National University of Defense Technology, 410073 Changsha, People's Republic of China*

*Hunan Key Laboratory of Intelligent Planning and Simulation for Aerospace Missions, 410073 Changsha, People's Republic of China*

and

Heng-nian Li‡

*Xi'an Satellite Control Center, 710043 Xi'an, People's Republic of China*

## I. Introduction

Optimization of orbit rendezvous is one of the basic technologies for the design and analysis of space missions [1]. When the magnitude of thrust is not very low, the orbit transfer can be approximated to a multi-impulse model. Many studies on impulsive orbit rendezvous of two-body model have been conducted [2-7]. In Ref. [2], the solution to Lambert's problem can be used to quickly obtain two-impulse transfer trajectories. For some particular orbits, an analytical solution of multiple impulses can be derived [3,4]. The optimization methods generally include indirect method using the prime vector theory [5] and direct method using parameter optimization [6,7]. Optimization of the two-body dynamics model does not take much time because the equations are analytical.

For the orbit rendezvous in low earth orbits (LEOs) such as debris removal missions, in-orbit service, and so on, the influence of perturbations must be considered. The main perturbations come from the $J_2$ term of the earth's non-spherical shape and the drag of atmosphere, which lead to a decrease in the semi-major axis and drift of the right ascension of the ascending node (RAAN) and argument of perigee [8,9]. When high-precision dynamics are involved in the trajectory optimization, the major difficulty is that the efficiency decreases significantly because of the time-consuming orbit prediction. When the transfer duration is short and the deviation caused by perturbations is small enough, the optimal trajectory of the two-body model can be applied as the initial value to shoot for the high-precision model. When the transfer duration is long, the deviation between the two models becomes too large for shooting methods. Taking advantage of the drift of orbit elements due to perturbations can greatly decrease the velocity increments required for rendezvous [10,11]. Most methods focus on directly searching for optimal impulses using evolutionary algorithms [12,13], and the perturbed Lambert's problem [14] can be adopted to accelerate the convergence. However, when the number of revolutions is hundreds or more, the effect of the same impulse on the neighboring revolutions are close but slightly different, which leads to many local optimal solutions. Hybrid method that adding the revolution number of impulses into the unknowns is studied in [15], but the mixed-encoding evolutionary algorithm is time-consuming.

---

* Engineer, State Key Laboratory of Astronautic Dynamics, also PhD Candidate, College of Aerospace Science and Engineering; hay04@foxmail.com.
† Corresponding author, Professor, College of Aerospace Science and Engineering, luoyz@nudt.edu.cn.
‡ Professor, State Key Laboratory of Astronautic Dynamics, henry_xscc@mail.xjtu.edu.cn

Based on the natural drift of the RAAN, Cerf [16] designed a strategy that used a drift orbit to reduce the velocity increment to eliminate the difference in the RAAN between the initial and target orbits. However, the optimal parameters of the drift orbit, such as the semi-major axis and inclination, were determined by traversal search. To obtain the optimal parameters of the drift orbit analytically, Huang et al. [17] established an equal constraint optimization model that could be quickly solved to obtain the approximated near-optimal velocity increment. However, as the analytical equation of the $J_2$ perturbation is used in the calculation, the solution is approximate and cannot be directly applied as the precise solution of full perturbations. Existing methods either applied simplified models to obtain approximate results or applied time-consuming evolutionary algorithms to obtain precise results. An efficient optimization method that can quickly obtain the precise impulses and trajectory is required for fast mission designing and trajectory generation especially when the count of targets is large.

The main contribution of this Note is that an iterative optimization method is proposed to quickly obtain the optimal high-precision trajectory by starting from an analytically estimated trajectory. Specifically, the optimal four impulses and transfer trajectory are simply expressed by ten parameters based on an existing estimation method from [17]. Prediction method of the orbital deviations between the chaser and target with high-precision dynamics is then proposed. Correction to the initial parameters is derived based on the analytical $J_2$ perturbed dynamics in a circular orbit. The prediction and correction processes are invoked repeatedly and form an iteration cycle to totally decrease the deviations and obtain the optimal impulses for high-precision rendezvous.

Simulation results proved that the iteration method adapts well to the full-perturbation dynamics and could always converge within five steps. Since only one orbit prediction is required in each iteration, the method is much more efficient than evolutionary algorithms while obtaining very close impulsive transfer trajectories.

## II. Problem description of Long Duration Perturbed Orbit Rendezvous

This problem can be described as follows. The initial orbital elements at $t_0$ of the chaser and target spacecraft are known. The chaser is required to rendezvous with the target at a given terminal time $t_f$ using multiple impulses. Thus, the optimization aims to find the impulses that satisfy the constraints on the orbital elements and minimize the fuel cost. In this Note, the orbit rendezvous in low-earth-orbits with a long transfer duration $\Delta t = t_f - t_0$ is studied.

The dynamics of a spacecraft can be expressed as:

$$\dot{\mathbf{r}} = \mathbf{v}$$
$$\mathbf{v} = -\frac{\mu}{r^3}\mathbf{r} + \mathbf{a}_{pert} \qquad (1)$$

where $\mathbf{r}$ and $\mathbf{v}$ are the position and velocity, which can be converted into and from orbital elements, $\mu$ is the gravity constant of the earth, $\mathbf{a}_{pert}$ is the perturbation, including the gravity of the sun and the moon, the earth's non-spherical perturbation, atmospheric drag, solar radial pressure, and so on.

For an impulsive orbit maneuver, the impulse is a vector of the velocity increment. Assume that the number of impulses is $n$, and the impulses are $\Delta \mathbf{v}_i, i = 1, 2, \ldots n$, then $\Delta \mathbf{v}_i$ can be directly added to the velocity as

$$\begin{aligned} \mathbf{r}(t_i^+) &= \mathbf{r}(t_i^-) \\ \mathbf{v}(t_i^+) &= \mathbf{v}(t_i^-) + \Delta \mathbf{v}_i \end{aligned} \quad (2)$$

where $t_i$ is the impulse time $\Delta \mathbf{v}_i$. The superscripts "+" and "-" mean the states after and before maneuver. Then, from the initial time, the states of the chaser at $t_f$ can be predicted using Eq. (1) and Eq. (2). Orbit rendezvous requires the following constraints:

$$\begin{aligned} \mathbf{r}(t_f) &= \mathbf{r}_f \\ \mathbf{v}(t_f) &= \mathbf{v}_f \end{aligned} \quad (3)$$

where $\mathbf{r}(t_f)$ and $\mathbf{v}(t_f)$ are the states of the chaser at $t_f$, and $\mathbf{r}_f$ and $\mathbf{v}_f$ are the states of the target at $t_f$. According to Chebyshev's equation of fuel and velocity increment, the minimal fuel is equal to the minimal velocity increment. Therefore, the objective function is written to minimize the total velocity increment:

$$J = \sum_{i=1}^{n} \left| \Delta \mathbf{v}_i \right| \quad (4)$$

To solve this problem, it is usually to assign and fix the number of impulses $n$ and use the three components of each $\Delta \mathbf{v}_i$ and the corresponding time $t_i$ as the parameters to be optimized. Evolutionary algorithms are always applied to obtain the optimal solution when the perturbed dynamics is considered in Eq. (1). However, when the duration is long, the time-consuming orbit prediction would greatly decrease the efficiency. Therefore, we provide an efficient iteration method to obtain the optimal rendezvous trajectory by starting with the estimated impulses in [17] as a good initial guess.

### III. Initial Value of Impulses and Prediction of Deviation

Ref. [17] designed a fast estimation model for the perturbed low earth orbit (LEO) rendezvous. In this section, the result is transformed to the initial values of impulses to calculate the deviation of the orbit elements between the chaser and target at $t_f$.

## A. Estimated Semi-Analytical Solution Considering $J_2$ perturbation

According to [17], we should first convert the osculating orbit elements into mean elements and calculate the difference between the orbit elements of the chaser and target without maneuvers. The difference can be expressed by a nonsingular orbit element:

$$\begin{aligned}
\Delta a_0 &= a(t_f) - a_f \\
\Delta e_{x0} &= e(t_f)\cos(\omega(t_f)) - e_f \cos(\omega_f) \\
\Delta e_{y0} &= e(t_f)\sin(\omega(t_f)) - e_f \sin(\omega_f) \\
\Delta i_0 &= i(t_f) - i_f \\
\Delta \Omega_0 &= \Omega(t_f) - \Omega_f \\
\Delta u_0 &= \omega(t_f) + f(t_f) - \omega_f - f_f
\end{aligned} \qquad (5)$$

where $[a(t_f), e(t_f), i(t_f), \Omega(t_f), \omega(t_f), f(t_f)]$ and $[a_f, e_f, i_f, \Omega_f, \omega_f, f_f]$ are the classical orbit elements of the chaser and target at $t_f$. Then, we assume that the optimal rendezvous strategy includes four impulses, two at the first revolution and two at the last revolution. The variation of the orbit elements for the two groups of impulses is then obtained by solving a semi-analytical optimization model. We denote $\Delta a_1$ and $\Delta a_2$ as the variation in the semi-major axis, $\Delta i_1$ and $\Delta i_2$ as the variation in the inclination, $\Delta \Omega_1$ and $\Delta \Omega_2$ as the variation of the RAAN, $k_1, k_2, k_3, k_4$ as the coefficients of eccentricity. The ten parameters can be solved analytically in [17] and the estimated total velocity increment can then be calculated.

The estimated value of each impulse and its corresponding time can also be determined using $[\Delta a_1, \Delta a_2, \Delta i_1, \Delta i_2, \Delta \Omega_1, \Delta \Omega_2, k_1, k_2, k_3, k_4]$. Assume that the four impulses are $\Delta \mathbf{v}_1, \Delta \mathbf{v}_2, \Delta \mathbf{v}_3$ and $\Delta \mathbf{v}_4$, which can be written as:

$$\begin{aligned}
\Delta \mathbf{v}_1 &= v_{t1}\mathbf{e}_t + v_{n1}\mathbf{e}_n + v_{r1}\mathbf{e}_r \\
\Delta \mathbf{v}_2 &= v_{t2}\mathbf{e}_t + v_{n2}\mathbf{e}_n + v_{r2}\mathbf{e}_r \\
\Delta \mathbf{v}_3 &= v_{t3}\mathbf{e}_t + v_{n3}\mathbf{e}_n + v_{r3}\mathbf{e}_r \\
\Delta \mathbf{v}_4 &= v_{t4}\mathbf{e}_t + v_{n4}\mathbf{e}_n + v_{r4}\mathbf{e}_r
\end{aligned} \qquad (6)$$

where $\mathbf{e}_t = \begin{bmatrix} v_x/v \\ v_y/v \\ v_z/v \end{bmatrix}, \mathbf{e}_r = \begin{bmatrix} x/r \\ y/r \\ z/r \end{bmatrix}, \mathbf{e}_n = \dfrac{\mathbf{e}_t \times \mathbf{e}_r}{|\mathbf{e}_t \times \mathbf{e}_r|}$ are the normalized vectors of the

tangential, radius, and normal directions. Additionally, $\mathbf{r} = [x, y, z]$ and $\mathbf{v} = [v_x, v_y, v_z]$ are the position and velocity, respectively, of the chaser. The three components of $\Delta\mathbf{v}_1, \Delta\mathbf{v}_2, \Delta\mathbf{v}_3$ and $\Delta\mathbf{v}_4$ are designed to satisfy the following constraints:

$$\left|\frac{v_{t1}}{v_{t2}}\right| = \left|\frac{v_{n1}}{v_{n2}}\right| = \left|\frac{v_{r1}}{v_{r2}}\right| \\ \left|\frac{v_{t3}}{v_{t4}}\right| = \left|\frac{v_{n3}}{v_{n4}}\right| = \left|\frac{v_{r3}}{v_{r4}}\right| \tag{7}$$

Then, according to the value of the ten parameters, the relationship between the components in $\Delta\mathbf{v}_1$ and $\Delta\mathbf{v}_2$ can be written as

$$\begin{aligned} v_{t1} + v_{t2} &= \frac{\Delta a_1}{2a_0}V \\ v_{t1} - v_{t2} &= \frac{k_1 V}{2} \\ v_{n1} - v_{n2} &= \sqrt{\Delta i_1^2 + (\Delta\Omega_1 \sin i_0)^2}\,V \\ v_{r1} - v_{r2} &= k_2 V \end{aligned} \tag{8}$$

where $a_0$ and $a_0$ are the initial semi-major axis and inclination of the chaser at the initial time, and $V$ is the mean velocity of the chaser at the initial time. Then, $\Delta\mathbf{v}_1$ and $\Delta\mathbf{v}_2$ can be solved using Eq. (7) and Eq. (8):

$$\begin{aligned} v_{t1} &= \frac{1}{2}(\frac{\Delta a}{2a} + \frac{k_1}{2})V \\ v_{t2} &= \frac{1}{2}(\frac{\Delta a}{2a} - \frac{k_1}{2})V \\ c_1 &= \left|\frac{v_{t1}}{v_{t2}}\right| \\ v_{n1} &= c_1\sqrt{\Delta i^2 + (\frac{\Delta\Omega \sin i}{2})^2}\,V \\ v_{n2} &= -(1-c_1)\sqrt{\Delta i^2 + (\frac{\Delta\Omega \sin i}{2})^2}\,V \\ v_{r1} &= c_1 k_3 V \\ v_{r2} &= -(1-c_1)k_3 V \end{aligned} \tag{9}$$

Similarly, the relationship between the components in $\Delta\mathbf{v}_3$ and $\Delta\mathbf{v}_4$ can be written as:

$$\begin{aligned}
v_{t3} + v_{t4} &= \frac{\Delta a_2}{2a_0} V \\
v_{t3} - v_{t4} &= \frac{k_2 V}{2} \\
v_{n3} - v_{n4} &= \sqrt{(\Delta i_2)^2 + (\Delta \Omega_2 \sin i_0)^2} V \\
v_{r3} - v_{r4} &= k_4 V
\end{aligned} \quad (10)$$

Then, $\Delta \mathbf{v}_3$ and $\Delta \mathbf{v}_4$ can be obtained as follows.

$$\begin{aligned}
v_{t3} &= \frac{1}{2}\left(\frac{\Delta a_2}{2a_0} + \frac{k_2}{2}\right)V \\
v_{t4} &= \frac{1}{2}\left(\frac{\Delta a_2}{2a_0} - \frac{k_2}{2}\right)V \\
c_2 &= \left|\frac{v_{t3}}{v_{t4}}\right| \\
v_{n3} &= c_2 \sqrt{(\Delta i_2)^2 + (\Delta \Omega_2 \sin i_0)^2} V \\
v_{n4} &= -(1-c_2)\sqrt{(\Delta i_2)^2 + (\Delta \Omega_2 \sin i_0)^2} V \\
v_{r3} &= c_2 k_4 V \\
v_{r4} &= -(1-c_2) k_4 V
\end{aligned} \quad (11)$$

The phase of each impulse can be calculated as:

$$\begin{aligned}
u_1 &= \arctan 2(\Delta \Omega_1 \sin i_0, \Delta i_1) \\
u_2 &= \arctan 2(\Delta \Omega_1 \sin i_0, \Delta i_1) + \pi \\
u_3 &= \arctan 2(\Delta \Omega_2 \sin i_0, \Delta i_2) \\
u_4 &= \arctan 2(\Delta \Omega_2 \sin i_0, \Delta i_2) + \pi
\end{aligned} \quad (12)$$

Thus, when $[\Delta a_1, \Delta a_2, \Delta i_1, \Delta i_2, \Delta \Omega_1, \Delta \Omega_2, k_1, k_2, k_3, k_4]$ is known, the details of $\Delta \mathbf{v}_1, \Delta \mathbf{v}_2, \Delta \mathbf{v}_3$ and $\Delta \mathbf{v}_4$ can also be obtained using Eq. (9), Eq. (11), and Eq. (12). The impulses have a one-to-one correspondence.

### B. Deviation of Estimated Solution

When $\Delta \mathbf{v}_1, \Delta \mathbf{v}_2, \Delta \mathbf{v}_3$, and $\Delta \mathbf{v}_4$ are known, the trajectory of the chaser and the deviation of the orbit elements at the rendezvous time can be calculated with Eq. (1) and Eq.(2) using numerical integration. The maneuvers are determined by the precise orbit and the phases $u_1, u_2, u_3, u_4$. In this Note, the orbits are considered to be near-circular orbits, so the true anomaly angle is almost equal to the mean anomaly angle. Then, the times of maneuvers

$t_1, t_2, t_3$, and $t_4$ can be calculated as follows:

First, we assume that the initial phase of the chaser is $u_0 = \omega_0 + M_0$. To ensure that $t_1$ is after $t_0$ and is as close as $t_0$, there are four conditions for the phase difference between $u_0$ and $u_1$:

$$\Delta u_1 = \begin{cases} u_1 - u_0, 0 < u_1 - u_0 < \pi \\ u_1 - u_0 - \pi, u_1 - u_0 > \pi \\ u_1 - u_0 + \pi, -\pi < u_1 - u_0 < 0 \\ u_1 - u_0 + 2\pi, u_1 - u_0 < -\pi \end{cases} \tag{13}$$

Eq. (13) ensures that the range of $\Delta u_1$ is $[0, \pi)$. Note that in the second and third conditions, $u_1$ and $u_2$ are exchanged, so the components of $\Delta \mathbf{v}_1$ and $\Delta \mathbf{v}_2$ should also be exchanged. Then, $t_1$ can be calculated as:

$$\Delta t_1 = t_1 - t_0 = \frac{\Delta u_1}{n_0} \tag{14}$$

where $n_0$ is the angular velocity of the initial orbit. The orbit of the chaser after $\Delta \mathbf{v}_1$ can then be obtained with Eq. (2) using the osculating position and velocity. In addition, $t_2$ can be calculated as:

$$\Delta t_2 = t_2 - t_1 = \frac{\pi}{n_1} \tag{15}$$

where $n_1$ is the mean angular velocity of the orbit after $\Delta \mathbf{v}_1$.

To ensure that $t_3$ and $t_4$ are on the last revolution, we can first calculate the phase of the chaser at $t_f$ by predicting the orbit after the second impulse at $t_2$. Then, the phase difference $\Delta u_3$ between the third maneuver and $u_f$ can be calculated as:

$$\Delta u_3 = \begin{cases} u_f - u_3, \pi < (u_f - u_3) < 2\pi \\ u_f - u_3 + \pi, 0 < (u_f - u_3) < \pi \\ u_f - u_3 + 2\pi, -\pi < (u_f - u_3) < 0 \\ u_f - u_3 + 3\pi, -2\pi < (u_f - u_3) < -\pi \end{cases} \tag{16}$$

where the range of $\Delta u_3$ is limited to $(2\pi, \pi]$. For the second and fourth conditions, $u_3$ and $u_4$ in Eq. (12) are exchanged, so the components of $\Delta \mathbf{v}_3$ and $\Delta \mathbf{v}_4$ should also be exchanged when the orbit is calculated. Then, we obtain the time of $\Delta \mathbf{v}_3$ as

$$t_3 = t_f - \frac{\Delta u_3}{n_2} \tag{17}$$

where $n_2$ is the mean angular velocity of the orbit after $\Delta \mathbf{v}_2$. In addition, after the third maneuver, $t_4$ can be obtained with Eq. (18).

$$t_4 = t_3 + \frac{\pi}{n_3} \tag{18}$$

where $n_3$ is the mean angular velocity of the orbit after $\Delta \mathbf{v}_3$. Then, the osculating orbit before and after the fourth maneuver can be obtained, and the orbit differences described by the non-singular orbital elements can be calculated using Eq. (19).

$$\begin{aligned}
\Delta a_p &= a_p - a_f \\
\Delta e_{xp} &= e_p \cos \omega_p - e_f \cos \omega_f \\
\Delta e_{yp} &= e_p \sin \omega_p - e_f \sin \omega_f \\
\Delta \Omega_p &= \Omega_p - \Omega_f \\
\Delta i_p &= i_p - i_f \\
\Delta u_p &= (\omega_p + f_p) - (\omega_f + f_f)
\end{aligned} \tag{19}$$

where $[a_p, e_p, i_p, \Omega_p, \omega_p, f_p]$ are converted to the mean orbital elements. $[\Delta a_p, \Delta e_{xp}, \Delta e_{yp}, \Delta i_p, \Delta \Omega_p, \Delta u_p]$ are the deviations of the orbital rendezvous when the estimated impulses are used.

### IV. Correction Algorithm and Iteration Process

To eliminate these deviations, an analytical modification method is proposed in this section based on the presupposed condition in which the orbits are nearly circular. The initial $[\Delta a_1, \Delta a_2, \Delta i_1, \Delta i_2, \Delta \Omega_1, \Delta \Omega_2, k_1, k_2, k_3, k_4]$ are defined as the parameters to be modified. Then, we can see that when $u_1, u_2, u_3, u_4$ are fixed: $\Delta u_p$ is mainly determined by $\Delta a_1$; $\Delta a_p$ is determined by $\Delta a_1$ and $\Delta a_2$; $\Delta i_p$ is determined by $\Delta i_1, \Delta i_2$; $\Delta \Omega_p$ is determined by

$\Delta a_1, \Delta i_1, \Delta \Omega_1, \Delta \Omega_2$; and $\Delta e_{xp}, \Delta e_{yp}$ are determined by, $\Delta a_1, \Delta a_2$, and $k_1, k_2, k_3, k_4$. Thus, we can eliminate the deviations of orbital elements sequentially.

## A. Correction to Semi-Major Axis

Assuming $\Delta a_1^c$ is the correction to $\Delta a_1$, the relationship between $\Delta a_1^c$ and $\Delta u_p$ is:

$$\Delta u_p = -\frac{3}{2}\frac{\Delta a_1^c n_0}{a_0 \Delta t} \Rightarrow \Delta a_1^c = -\frac{2\Delta u_p a_0 \Delta t}{3 n_0} \tag{20}$$

where $\Delta t = t_f - t_0$ is the drift duration and $n_0$ is the initial angular velocity. Then, $\Delta a_2$ should be corrected to:

$$\Delta a_2 + \Delta a_2^c = \Delta a_2 - \Delta a_1^c - \Delta a_p \tag{21}$$

## B. Correction to Inclination and RAAN

Assuming $\Delta i_1^c$ and $\Delta i_2^c$ are the correction to $\Delta i_1$ and $\Delta i_2$, respectively. Because we don't correct the phases of the impulses, the correction to $\Delta \Omega_1$ and $\Delta \Omega_2$ can be calculated as

$$\frac{\Delta \Omega_1 + \Delta \Omega_1^c}{\Delta i_1 + \Delta i_1^c} = \frac{\Delta \Omega_1}{\Delta i_1} = \frac{\tan u_1}{\sin i_0}$$
$$\frac{\Delta \Omega_2 + \Delta \Omega_2^c}{\Delta i_2 + \Delta i_2^c} = \frac{\Delta \Omega_2}{\Delta i_2} = \frac{\tan u_3}{\sin i_0} \tag{22}$$

The change in the RAAN also includes the drift change by $\Delta i_1^c$ and $\Delta a_1^c$. To eliminate the deviation of inclination and the RAAN, a two-dimensional equation is obtained as

$$(\Delta i_1 + \Delta i_1^c) + (\Delta i_2 + \Delta i_2^c) = \Delta i_0 - \Delta i_p$$
$$(\Delta i_1 + \Delta i_1^c)\frac{\tan u_1}{\sin i_0} + (\Delta i_2 + \Delta i_2^c)\frac{\tan u_3}{\sin i_0} + (\dot{\Omega} + \Delta \dot{\Omega})\Delta t = \Delta \Omega_0 - \Delta \Omega_p \tag{23}$$

where $\Delta \dot{\Omega}$ is the variation in RAAN drift after the first and second impulses:

$$\Delta \dot{\Omega} \approx \dot{\Omega}(-3.5\frac{\Delta a_1}{a_0} - \tan i_0 \Delta i_1) \tag{24}$$

Then, Eq. (23) can be rewritten as:

$$\Delta i_1^c + \Delta i_2^c = -\Delta i_p$$
$$\frac{\tan u_1}{\sin i_0}\Delta i_1^c + \frac{\tan u_3}{\sin i_0}\Delta i_2^c + \dot{\Omega}(-3.5\frac{\Delta a_1^c}{a_0} - \Delta i_1^c \tan i_0)\Delta t = -\Delta \Omega_p \quad (25)$$

The solution of this linear equation is:

$$\Delta i_1^c = \frac{-\dfrac{\tan u_3 \Delta i_p}{\sin i_0} + \Delta \Omega_p + \dot{\Omega}(-3.5\dfrac{\Delta a_1^c}{a_0})\Delta t}{\dfrac{\tan u_3}{\sin i_0} - \dfrac{\tan u_1}{\sin i_0} + \dot{\Omega}\tan i_0 \Delta t} \quad (26)$$
$$\Delta i_2^c = -\Delta i_1^c - \Delta i_p$$

Then, $\Delta \Omega_1^c$ and $\Delta \Omega_2^c$ can be obtained using Eq. (22).

**C. Correction to Eccentricity**

In the optimization model in Ref. [17], Eqs. (27) is used to obtain the optimal values of $k_1, k_2, k_3$, and $k_4$. $k_1$ and $k_2$ are, respectively, the eccentricity changed by the tangential component of the front two impulses and the last two impulses, while $k_3$ and $k_4$ are, the eccentricity changed by the radius components.

$$\Delta e_{x0} = k_1 \frac{\Delta a_1}{a_0}\cos(u_1 + \Delta u_1) + k_2 \frac{\Delta a_2}{a_0}\cos u_3 + k_3 \sin(u_1 + \Delta u_1) + k_4 \sin u_3$$
$$\Delta e_{y0} = k_1 \frac{\Delta a_1}{a_0}\sin(u_1 + \Delta u_1) + k_2 \frac{\Delta a_2}{a_0}\sin u_3 + k_3 \cos(u_1 + \Delta u_1) + k_4 \cos u_3 \quad (27)$$

where $\Delta u_1$ is the drift of $\omega$ caused by perturbation:

$$\Delta u_1 = -1.5 J_2 \left(\frac{R_e}{a + \Delta a + \Delta a_1}\right)^2 (5\cos^2(i_0 + \Delta i + \Delta i_1) - 1)\Delta t \quad (28)$$

Assuming $\Delta k_1, \Delta k_2, \Delta k_3, \Delta k_4$ are the correction values of $k_1, k_2, k_3$, and $k_4$. Then, the corresponding variation in eccentricity should be equal to the deviation in Eq. (19). Thus, we can obtain:

$$\Delta e_{x0} - \Delta e_{xp} = (k_1 + \Delta k_1)\frac{\Delta a + \Delta a_1}{a_0}\cos(u_1 + \Delta u_1) + (k_2 + \Delta k_2)\frac{\Delta a_0 - \Delta a_p - \Delta a - \Delta a_1}{a_0}\cos u_3$$
$$+(k_3 + \Delta k_3)\sin(u_1 + \Delta u_1) + (k_4 + \Delta k_4)\sin u_3 \quad (29)$$
$$\Delta e_{y0} - \Delta e_{yp} = (k_1 + \Delta k_1)\frac{\Delta a + \Delta a_1}{a_0}\sin(u_1 + \Delta u_1) + (k_2 + \Delta k_2)\frac{\Delta a_0 - \Delta a_p - \Delta a - \Delta a_1}{a_0}\sin u_3$$
$$+(k_3 + \Delta k_3)\cos(u_1 + \Delta u_1) + (k_4 + \Delta k_4)\cos u_3$$

which can be simplified as:

$$-\Delta e_{xp} = k_1 \frac{\Delta a_1}{a_0}\cos(u_1+\Delta u_1) + \Delta k_1 \frac{\Delta a + \Delta a_1}{a_0}\cos(u_1+\Delta u_1) + k_2 \frac{-\Delta a_p - \Delta a_1}{a_0}\cos u_3$$
$$+\Delta k_2 \frac{\Delta a_0 - \Delta a_p - \Delta a - \Delta a_1}{a_0}\cos u_3 + \Delta k_3 \sin(u_1+\Delta u_1) + \Delta k_4 \sin u_3$$
$$-\Delta e_{xp} = k_1 \frac{\Delta a_1}{a_0}\sin(u_1+\Delta u_1) + \Delta k_1 \frac{\Delta a + \Delta a_1}{a_0}\sin(u_1+\Delta u_1) + k_2 \frac{-\Delta a_p - \Delta a_1}{a_0}\sin u_3$$
$$+\Delta k_2 \frac{\Delta a_0 - \Delta a_p - \Delta a - \Delta a_1}{a_0}\sin u_3 + \Delta k_4 \cos(u_1+\Delta u_1) + \Delta k_4 \cos u_3 \qquad (30)$$

The optimal $\Delta k_1, \Delta k_2, \Delta k_3, \Delta k_4$ should ensure that Eq. (30) is satisfied and that the total velocity increment is minimized. This is also an equal-constraint optimization:

$$\min f = |\Delta v_1| + |\Delta v_2| + |\Delta v_3| + |\Delta v_4|$$

$$|\Delta v_1| + |\Delta v_2| = \begin{cases} V\sqrt{(\frac{\Delta a_1 + \Delta a_1^c}{2a_0})^2 + (\Delta i_1 + \Delta i_1^c)^2 + ((\Delta \Omega_1 + \Delta \Omega_1^c)\sin i_0)^2 + (k_3 + \Delta k_3)^2}, & |k_1 + \Delta k_1| \le \left|\frac{\Delta a_1 + \Delta a_1^c}{a_0}\right| \\ V\sqrt{(\frac{k_1 + \Delta k_1}{2})^2 + (\Delta i_1 + \Delta i_1^c)^2 + ((\Delta \Omega_1 + \Delta \Omega_1^c)\sin i_0)^2 + k_3^2(k_3 + \Delta k_3)^2}, & |k_1 + \Delta k_1| > \left|\frac{\Delta a_1 + \Delta a_1^c}{a_0}\right| \end{cases}$$

$$|\Delta v_3| + |\Delta v_4| = \begin{cases} V\sqrt{(\frac{\Delta a_2 + \Delta a_2^c}{2a_0})^2 + (\Delta i_2 + \Delta i_2^c)^2 + ((\Delta \Omega_2 + \Delta \Omega_2^c)\sin i_0)^2 + (k_4 + \Delta k_4)^2}, & |k_2 + \Delta k_2| \le \left|\frac{\Delta a_2 + \Delta a_2^c}{a_0}\right| \\ V\sqrt{(\frac{k_2 + \Delta k_2}{2})^2 + (\Delta i_2 + \Delta i_2^c)^2 + ((\Delta \Omega_2 + \Delta \Omega_2^c)\sin i_0)^2 + (k_4 + \Delta k_4)^2}, & |k_2 + \Delta k_2| > \left|\frac{\Delta a_2 + \Delta a_2^c}{a_0}\right| \end{cases}$$

(31)

Eq. (31) can be solved using the minimal principle, which is similar to the algorithm in [17]. However, in this Note, the differential evolution (DE) algorithm [18] is adopted. Note that when $\Delta k_1$ and $\Delta k_2$ are known, $\Delta k_3$ and $\Delta k_4$ can be directly expressed by $\Delta k_1$ and $\Delta k_2$:

$$\Delta k_4 = \frac{B_1 - B_2 \tan(u_1 + \Delta u_1)}{\sin u_3 - \cos u_3 \tan(u_1 + \Delta u_1)}$$
$$\Delta k_3 = \frac{B_1 - \Delta k_4 \sin u_3}{\sin(u_1 + \Delta u_1)}$$
$$B_1 = -\Delta e_{xp} - k_1 \frac{\Delta a_1}{a_0}\cos(u_1+\Delta u_1) - k_2 \frac{-\Delta a_p - \Delta a_1}{a_0}\cos u_3$$
$$-\Delta k_1 \frac{\Delta a + \Delta a_1}{a_0}\cos(u_1+\Delta u_1) - \Delta k_2 \frac{\Delta a_0 - \Delta a_p - \Delta a - \Delta a_1}{a_0}\cos u_3 \qquad (32)$$
$$B_2 = -\Delta e_{xp} - k_1 \frac{\Delta a_1}{a_0}\sin(u_1+\Delta u_1) - k_2 \frac{-\Delta a_p - \Delta a_1}{a_0}\sin u_3$$
$$-\Delta k_1 \frac{\Delta a + \Delta a_1}{a_0}\sin(u_1+\Delta u_1) - \Delta k_2 \frac{\Delta a_0 - \Delta a_p - \Delta a - \Delta a_1}{a_0}\sin u_3$$

Then, the dimension of the unknowns can be reduced to two, and Eq. (31) becomes an equal constraint optimization. DE is more time-consuming than analytical solving methods but is still acceptable to solve Eq. (31) because the unknowns are only two-dimensional and the calculation is much less than that of a long duration orbit prediction via numerical integration.

**D. Iteration Process**

Note that the correction method is based on the analytical dynamic equations of circular orbit, so, the parameters and the impulses after one correction is still approximate. To further decrease the deviation of the rendezvous, an iteration algorithm can be designed. The flowchart is shown in Algorithm 1. In each step of the iteration, when $\Delta a_1^c$, $\Delta a_2^c$, $\Delta i_1^c$, $\Delta i_2^c$, $\Delta \Omega_1^c$, $\Delta \Omega_2^c$, $\Delta k_1, \Delta k_2, \Delta k_3$ and $\Delta k_4$ are obtained, the corresponding impulses can be calculated using Eq. (12) and Eq. (13). Then, the orbit deviations of the rendezvous can be updated using Eq. (1) and Eq. (2). The correction to the control parameters can also be updated. After several steps to convergent, the optimal impulsive transfer trajectory can be obtained.

**Algorithm 1: Iteration method to eliminate orbital deviations**

| | |
|---|---|
| 1 | Obtain initial ten parameters: $[\Delta a_1, \Delta a_2, \Delta i_1, \Delta i_2, \Delta \Omega_1, \Delta \Omega_2, k_1, k_2, k_3, k_4]$ |
| 2 | Transform the parameters to phases and magnitudes of four impulses |
| 3 | Predict the orbit and calculate the deviation with target: $[\Delta a_p, \Delta e_{xp}, \Delta e_{yp}, \Delta i_p, \Delta \Omega_p, \Delta u_p]$ |
| 4 | **while** deviations don't meet requirement **do** |
| 5 | $\quad$ Calculate correction to $\Delta a_1$ by Eq. (20) |
| 6 | $\quad$ Calculate correction to $\Delta a_2$ by Eq. (21) |
| 7 | $\quad$ Calculate corrections to $\Delta i_1, \Delta i_2, \Delta \Omega_1$ and $\Delta \Omega_2$ by Eqs. (26) and (22) |
| 8 | $\quad$ Re-optimize corrections to $k_1, k_2, k_3$ and $k_4$ by DE |
| 9 | $\quad$ Update phases and magnitudes of four impulses |
| 10 | $\quad$ Predict the orbit and update $[\Delta a_p, \Delta e_{xp}, \Delta e_{yp}, \Delta i_p, \Delta \Omega_p, \Delta u_p]$ |
| 11 | **end while** |
| 12 | Output optimal impulses and precise trajectory |

## V. Simulation Results

In this Note, two types of orbit rendezvous in LEO with different altitudes and inclinations were tested. One was the debris removal mission in the solar synchronous orbit (SSO), and the other was the Automated Transfer Vehicle (ATV) mission, which was highly influenced by the drag of the atmosphere. Perturbations due to non-spherical gravity (20 × 20), the gravity of the sun and the moon, drag, and solar radial pressure are all considered in Eq. (1).

**A. Debris Removal Mission**

The 9th global trajectory optimization competition provided an optimization problem for multiple debris removal mission using multiple spacecraft, which has attracted a great number of teams to participate in. To invalidate the method in this Note, a transfer segment from the champion's results in [19] is selected. The orbit elements of the chaser and the target are listed in Table 1. The chaser starts at $t_0$ =0 days and should rendezvous with the target at $t_f$ =24.88 days. The chaser area-mass ratio was set to 0.01. In the calculation, it was found that adding a coefficient $k_e < 1$ to $\Delta e_{xp}$ and $\Delta e_{yp}$ could significantly accelerate convergence. This is because the corrections are calculated based on the circular orbit assumption; however, there may be a combined influence of the tangential impulse and the radial impulse on the eccentricity and phase. Therefore, the deviation in eccentricity cannot be completely removed. The convergence of the iteration is shown in Table 2. Finally, the total velocity increment is 161.94 m/s.

**Table 1 Detail of orbital elements at $t_0$**

|        |            | $a$ (m)    | $e$         | $i$ (deg) | $\Omega$ (deg) | $\omega$ (deg) | $M$ (deg) |
|--------|------------|------------|-------------|-----------|----------------|----------------|-----------|
| Chaser | Mean       | 7157397.82 | 0.015212152 | 98.6435   | 137.6365       | 64.1761        | 322.3407  |
|        | Osculating | 7163168.70 | 0.015493283 | 98.6390   | 137.6336       | 65.0002        | 321.5570  |
| Target | Mean       | 7111954.45 | 0.0072194   | 97.4512   | 138.0585       | 91.3079        | 23.7845   |
|        | Osculating | 7106032.90 | 0.006229437 | 97.4530   | 138.0640       | 83.4400        | 31.6064   |

**Table 2 Orbit deviations during the iteration**

| Iteration | $\Delta a_p$ (m) | $\Delta e_{xp}$ | $\Delta e_{yp}$ | $\Delta i_p$ (deg) | $\Delta \Omega_p$ (deg) | $\Delta u_p$ (deg) |
|-----------|------------------|-----------------|-----------------|--------------------|-------------------------|--------------------|
| 0 | -723.2258 | -8.8966E-05 | 1.0294E-04 | -2.9043E-03 | -2.0875E-02 | 1.4491E+01 |
| 1 | -15.9695  | -4.1173E-05 | 6.1210E-05 | 1.2690E-04  | 3.8323E-03  | -8.9788E-03 |
| 2 | -2.7941   | -6.8882E-05 | 7.8459E-05 | -4.5655E-05 | -5.9899E-04 | 4.6099E-02 |
| 3 | 1.8042    | -5.4662E-05 | 6.9991E-05 | 1.8421E-05  | 1.2516E-04  | -3.9502E-02 |
| 4 | -0.9687   | -6.1867E-05 | 7.4253E-05 | -8.7955E-06 | -4.0134E-05 | 2.2015E-02 |
| 5 | 0.4971    | -5.8227E-05 | 7.2095E-05 | 4.3820E-06  | 1.7295E-05  | -1.1403E-02 |

**Table 3 Detail of impulses during the iteration**

| Iteration | $t_1$ (day) | $\Delta v_1$ (m/s) | $t_2$ (day) | $\Delta v_2$ (m/s) | $t_3$ (day) | $\Delta v_3$ (m/s) | $t_4$ (day) | $\Delta v_4$ (m/s) | Sum (m/s) |
|-----------|-------------|--------------------|-------------|--------------------|-------------|--------------------|-------------|--------------------|-----------|
| 0 | 0.0300 | 28.242 | 0.0649 | 99.562  | 24.8462 | 1.351 | 24.8809 | 32.582 | 161.8159 |
| 1 | 0.0300 | 29.244 | 0.0650 | 100.258 | 24.8439 | 0.785 | 24.8786 | 31.831 | 162.0128 |
| 2 | 0.0300 | 28.987 | 0.0650 | 100.243 | 24.8440 | 0.758 | 24.8787 | 32.031 | 161.9159 |
| 3 | 0.0300 | 29.085 | 0.0650 | 100.203 | 24.8439 | 0.786 | 24.8787 | 31.985 | 161.9542 |
| 4 | 0.0300 | 29.041 | 0.0650 | 100.229 | 24.8440 | 0.768 | 24.8787 | 32.002 | 161.9357 |
| 5 | 0.0300 | 29.063 | 0.0650 | 100.215 | 24.8439 | 0.778 | 24.8787 | 31.994 | 161.9449 |

It can be observed that the final eccentricity error is less than 1e-4, and the phase error is less than 0.02° after five iterations. The errors of the other orbit elements were close to zero. The position error is less than 1.5 km, which can easily satisfy the distance requirement for the chaser to switch to autonomous control using relative dynamics. The variations in the mean orbit elements are shown in Figs. 2 and 3.

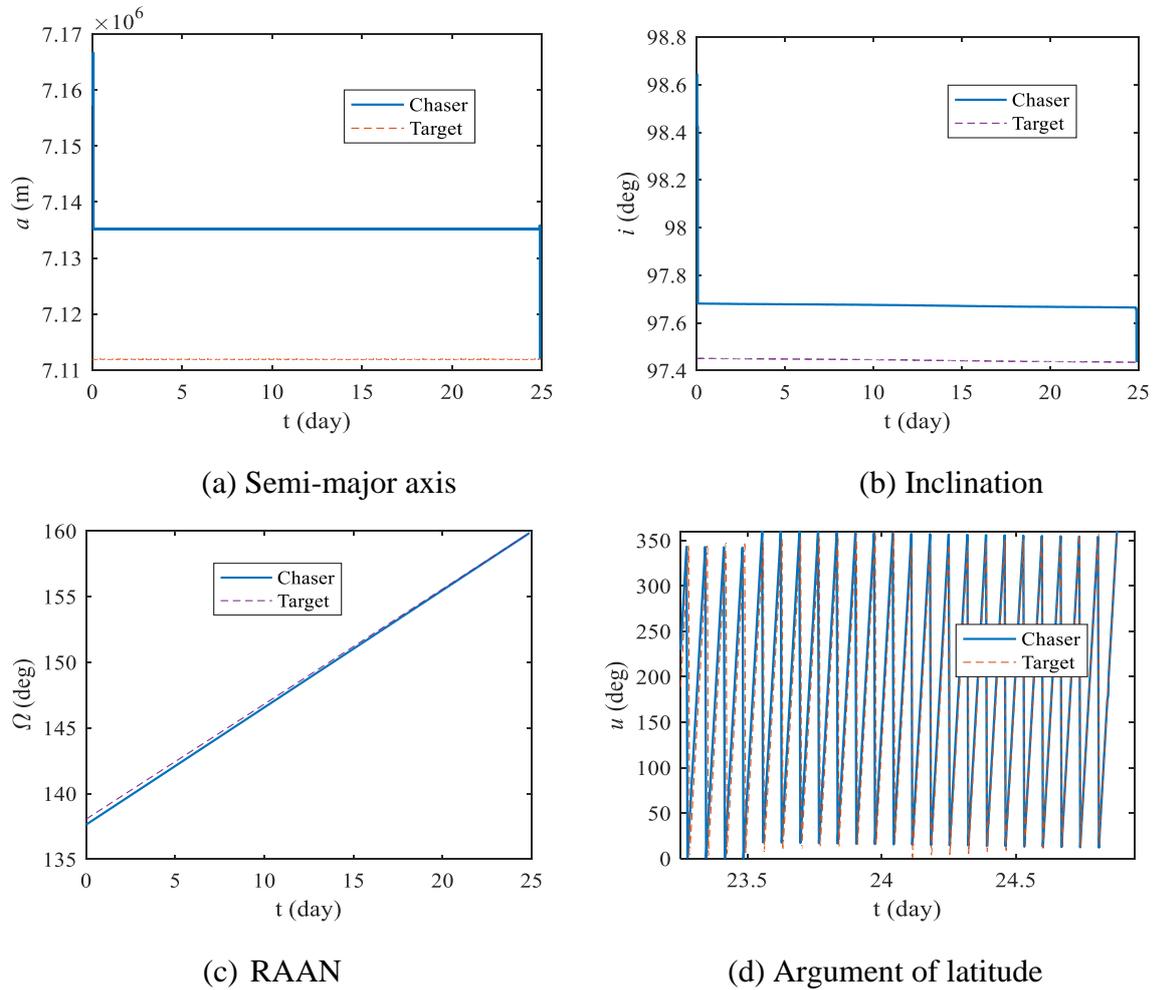

(a) Semi-major axis    (b) Inclination

(c) RAAN    (d) Argument of latitude

**Fig. 2 Histories of** $a, i, \Omega$, **and** $u$.

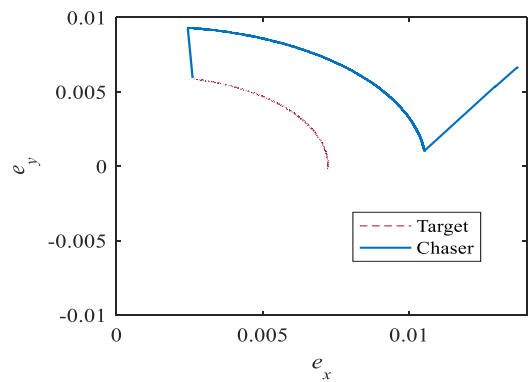

**Fig. 3 History of** $e_x$ **and** $e_y$.

If higher precision is required, the perturbed Lambert's problem can be used to recalculate the third and fourth impulses. The solution to Lambert's problem can be written as a function:

$$[\mathbf{v}^{in}, \mathbf{v}^{out}] = \text{Lambert}(\mathbf{r}^{in}, \mathbf{r}^{out}, \Delta t) \tag{33}$$

where $\mathbf{r}^{in}$ and $\mathbf{r}^{out}$ are the given initial position and target position, $\Delta t$ is the given time of the arc, $\mathbf{v}^{in}$ and $\mathbf{v}^{out}$ are the solutions, which are the departing velocity and the arriving velocity, respectively. We can set the position of chaser at $t_3$ as the input position vector and set the position of the target at $t_4$ as the shooting target. Then, $\Delta \mathbf{v}_3$ and $\Delta \mathbf{v}_4$ can be calculated as:

$$\begin{aligned} \Delta \mathbf{v}_3 &= \mathbf{v}^{in} - \mathbf{v}(t_3^-) \\ \Delta \mathbf{v}_4 &= \mathbf{v}(t_4^+) - \mathbf{v}^{out} \end{aligned} \tag{34}$$

where $\mathbf{v}(t_3^-)$ is the velocity of the chaser before $\Delta \mathbf{v}_3$, and $\mathbf{v}(t_4^+)$ is the velocity of the target at $t_4$. In this Note, to overcome the singularity problem when the arc is close to $\pi$, the normal components of $\mathbf{v}^{in}$ are fixed to the result after the iteration in Section IV.D. The perturbed Lambert's problem only needs to recalculate the tangential and radial components. To obtain the precise $\Delta \mathbf{v}_3$ and $\Delta \mathbf{v}_4$, the numerical orbit propagation calculation also needs to be called several times. However, the duration is only half of the orbital period, so this calculation does not require much time. After this step, the deviation in the position could be less than 30 m.

The same transfer can be seen in the second chain in [19] and the optimal velocity increment is 161.8 m/s after more than 2000 iterations by SNOPT (an SQP algorithm for large-scale constrained optimization). Each revolution has been discretized to 10 points to accelerate the convergence in their method. By contrast, differential evolutionary (DE) algorithm is also used to directly solve the optimization model in [13] and obtain a result of 163.6 m/s after 80000 generations. Note that only $J_2$ perturbation is considered in the comparison because the full dynamics orbit prediction is too time-consuming for DE to convergent.

After a test of different transfers in the submissions from [19] and [20], it's seen that the total velocity increments by the iteration method were slightly larger than those of the evolutionary algorithm (the mean relative error is less than 1.1%). The comparison of efficiency is detailed in Table 4. It's seen that based on the initial solutions obtained by estimation method in [17], the iteration method in this Note can obtain the optimal trajectory much more quickly than previous optimization methods. The iteration always converged within five orbit predictions and corrections.

**Table 4 Details of comparison**

| Method | Number of orbit predictions |
|---|---|
| Iteration method | 5 |
| Estimation method in [17] | 0 |
| Method in [19] | 2000 |
| Evolutionary optimizations in [13] and [20] | 80000 |

**B. Automated Transfer Vehicle mission**

The problem is a practical fourteen-day (two-week) rendezvous phasing mission [12]. The initial orbits are listed in Table 5, and the parameters of the force model are the same as those in Ref. [12]. Because the altitude is quite low, the drag of the atmosphere is much stronger than that in Section 4.1. To be consistent with the constraint on time in [12], the revolution number of second impulse is set to 18 and revolution number of third impulse is 195. Then, the middle time of the front two impulses is approximately equal to $t_0 + 9T$, and the middle time of the last two impulses is approximately equal to $t_f - 9T$, where $T$ is the period of initial orbit. So, when calculating the parameters $[\Delta a_1, \Delta a_2, \Delta i_1, \Delta i_2, \Delta \Omega_1, \Delta \Omega_2, k_1, k_2, k_3, k_4]$ by estimation method in [17], the transfer duration should be approximately equal to $t_f - t_0 - 18T$. Moreover, Eq. (15), Eq. (17) and Eq. (18) should be modified to Eq. (35). Then the iteration process can be applied to obtain the transfer trajectory.

$$\begin{aligned}
\Delta t_2 &= t_2 - t_1 = \frac{\pi + 34\pi}{n_1} \\
t_3 &= t_f - \frac{\Delta u_3 - 34\pi}{n_2} \\
t_4 &= t_3 + \frac{\pi + 34\pi}{n_3}
\end{aligned} \quad (35)$$

**Table 5 Details of orbit elements**

| | | $a$ (m) | $e$ | $i$ (deg) | $\Omega$ (deg) | $\omega$ (deg) | $M$ (deg) |
|---|---|---|---|---|---|---|---|
| Chaser | Mean | 6640200.281 | 0.008816 | 42.062 | 171.630 | 121.178 | 359.801 |
| | Osculating | 6638140 | 0.009039 | 42.05 | 171.6 | 120 | 0.982 |
| Target | Mean | 6722943.328 | 0.000226 | 42.015 | 169.177 | 114.469 | 130.520 |

|          | Osculating | 6720140 | 1e-5 | 42 | 169.2 | 100 | 145 |

The orbit deviations during the iterations are listed in Table 6. The chaser finally needed 62.6 m/s to complete the rendezvous, which is less than the result in [12]. The mean semi-major axis is shown in Fig. 4, which is close to the Fig.3 in [12]. It can be seen the iteration corrected the different decreasing rates of semi-major axis caused by the atmosphere.

**Table 6 Orbit deviations during the iteration**

| Iteration | $\Delta a_p$ (m) | $\Delta e_{xp}$ | $\Delta e_{yp}$ | $\Delta i_p$ (deg) | $\Delta \Omega_p$ (deg) | $\Delta u_p$ (deg) |
|---|---|---|---|---|---|---|
| 0 | 5404.696 | 1.515E-05 | -6.785E-04 | -4.191E-03 | 1.441E-01 | -5.656E+01 |
| 1 | -555.747 | 3.906E-05 | -4.292E-04 | -9.769E-05 | -1.905E-02 | 7.242E+00 |
| 2 | 64.07027 | 2.685E-05 | -5.418E-04 | 3.645E-05 | 1.235E-03 | -4.572E-01 |
| 3 | -2.04826 | 3.210E-05 | -4.873E-04 | -2.797E-06 | -4.114E-05 | 1.896E-02 |
| 4 | -1.23728 | 2.984E-05 | -5.145E-04 | 8.165E-07 | -1.686E-05 | 4.000E-03 |

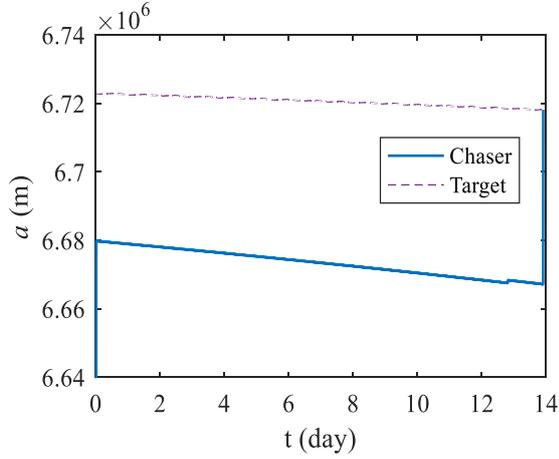
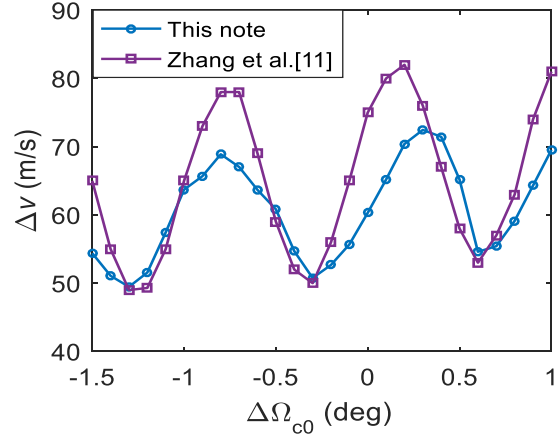

**Fig. 4 Variation of mean semi-major axis. Fig. 5 Comparison of different initial RAAN.**

In addition, to compare with the optimization results in Ref. [12] having different RAAN differences, we added $\Delta \Omega_{c0} \in [-1.5, 1.0]$ to the chaser's initial RAAN and obtain the optimal impulses. The results and comparisons are presented in Fig. 5. Most of the velocity increments are less than those in Ref. [12]. This is because the tangential impulse and normal impulse of the maneuver strategy in Ref. [12] are separate. Therefore, the total velocity increment is an absolute value sum of the different directions. The result in this Note is more like a quadratic sum, and the iteration process is much simpler.

## VI.  Conclusion

A novel fast optimization method for long-duration perturbed orbit rendezvous is proposed in this Note. Firstly, the optimal trajectory is expressed by ten parameters which can be analytically estimated as the initial values. Then, the orbital deviations at the rendezvous time can be predicted and an analytical correction to the impulses can be approximately obtained based on the $J_2$ perturbed dynamics equation of circular orbits. Iteration process is then designed to quickly obtain a high-precision solution by repeating deviations prediction and parameters corrections. The simulation results prove that the solution is remarkably close to that optimized by evolutionary algorithms, and the calculation is much more efficient. The iterative method in this Note can be well applied in the mission analysis and trajectory optimization of long-duration orbit rendezvous in low earth orbits.

## Funding Sources

This work was supported by the National Natural Science Foundation of China (No. 11972044).

## References

[1] Abolfazl, S., Josu, C., and Jose, A. L., "Spacecraft trajectory optimization: A review of models, objectives, approaches and solutions," *Progress in Aerospace Sciences*, Vol. 102, 2018, pp. 76-98.
https://doi.org/10.2514/1.G001382

[2] Arlulkar, P. V., and Naik, S. D., "Solution based on dynamical approach for multiple-revolution lambert problem," Journal of Guidance, Control, and Dynamics, Vol. 34, No. 3, 2011, pp. 920–923.
https://doi.org/10.2514/1.51723

[3] Prussing, J. E., "Optimal two- and three-impulse fixed-time rendezvous in the vicinity of a circular orbit," *AIAA Journal*, Vol. 8, No. 7, 1970, pp. 1221–1228.
https://doi.org/10.2514/3.5876

[4] Prussing, J. E., and Chiu, J. H., "Optimal multiple-impulse time-fixed rendezvous between circular orbits," *Journal of Guidance, Control, and Dynamics*, Vol. 9, No. 1, 1986, pp. 17–22.
https://doi.org/10.2514/3.20060

[5] Handelsman, M., and Lion, P. M., "Primer vector on fixed-time impulsive trajectories," *AIAA Journal*, Vol. 6, No. 1, 1968, pp. 127–135.
https://doi.org/10.2514/3.4452

[6] Riggi, L., and D'Amico, S., "Optimal impulsive closed-form control for spacecraft formation flying and rendezvous," *2016 American Control Conference (ACC),* Boston, MA, USA, 2016, pp. 5854–5861.
https://doi.org/10.1109/ACC.2016.7526587

[7] Fossa, A., Bettanini, C., "Optimal rendezvous trajectory between Sample Return Orbiter and Orbiting Sample Container in a Mars Sample Return mission," *Acta Astronautica*, Vol.171, 2020, pp.31–41.
https://doi.org/10.1016/j.actaastro.2020.02.046

[8] Gurfil, P., "Analysis of $J_2$-perturbed motion using mean non-osculating orbital elements," *Celestial Mechanics &*

*Dynamical Astronomy*, Vol. 90, No. 3, 2004, pp. 289–306.

https://doi.org/10.1007/s10569-004-0890-x

[9] Vallado, D. A., *Fundamentals of Astrodynamics and Applications*, 2nd ed., Microscosm Press, El Segundo, CA, 2001, pp. 644–652.

[10] Ocampo, C., Guinn, J., and Breeden, J., "Rendezvous options and dynamics for the mars sample return mission," *Advances in the Astronautical Sciences*, Vol. 109, Univelt, San Diego, CA, 2002, pp. 1661–1680.

[11] Labourdette, P., and Baranov, A. A., "A software for rendezvous between near-circular orbits with large initial ascending node difference," *Proceedings of the 17th International Symposium on Space Flight Dynamics*, Keldysh Inst. of Applied Mathematics, Russian Academy of Sciences Paper ISSFD-2003-1105, Moscow, June 2003.

[12] Zhang, J., Tang, G. J., and Luo, Y. Z., "Optimization of an orbital long-duration rendezvous mission," *Aerospace Science and Technology*, Vol. 58, 2016, pp. 482–489.

https://doi.org/10.1016/j.ast.2016.09.011

[13] Casalino, L. and Dario, P., "Active debris removal missions with multiple targets," *AIAA/AAS Astrodynamics Specialist Conference*. AIAA Paper 2014–4226, Aug. 2014.

https://doi.org/10.2514/6.2014-4226

[14] Zhang, G., Mortari, D., and Zhou, D., "Constrained multiple-revolution lambert's problem," *Journal of Guidance, Control, and Dynamics*, Vol. 33, No. 6, 2010, pp. 1779–1786.

https://doi.org/10.2514/6.2014-4226

[15] Zhang, J., Wang, X., Ma, X.B., Yi, T., and Huang, H.B., "Spacecraft long-duration phasing maneuver optimization using hybrid approach," *Acta Astronautica*, Vol. 72, 2012, pp. 132–142.

https://doi.org/10.1016/j.actaastro.2011.09.008

[16] Cerf, M., "Multiple space debris collecting mission: optimal mission planning," *Journal of Optimization Theory & Applications*, Vol. 167, No. 1, 2015, pp. 195–218.

https://doi.org/10.1007/s10957-015-0705-0

[17] Huang, A. Y., Luo, Y. Z., and Li, H. N., "Fast estimation of perturbed impulsive rendezvous via semi-analytical equality-constrained optimization," *Journal of Guidance, Control, and Dynamics*, Vol. 43, No. 12, 2020, pp. 2383–2390.

https://doi.org/10.2514/1.G005220

[18] Price, K., Storn, R., and Lampinen, J., *Differential Evolution—A Practical Approach to Global Optimization*. Berlin, Germany: Springer, 2005.

[19] Petropoulos, A. E., Grebow, D., Jones, D., Lantoine, G., Nicholas, A., Roa, J., Senent, J., Stuart, J., Arora, N., Pavlak, T., Lam, T., McElrath, T., Roncoli, R., Garza, D., Bradley, N., Landau, D., Tarzi, Z., Laipert, F., Bonfiglio, E., Wallace, M., and Sims, J., "GTOC9: Results from the Jet Propulsion Laboratory (Team JPL)," *Acta Futura*, Vol. 11, 2018, pp. 25–35.

https://doi.org/10.5281/zenodo.1139152

[20] Chen, S. Y., Li, H. Y., Jiang, F. H., Baoyin, H., "Optimization for Multitarget, Multispacecraft Impulsive Rendezvous Considering J2 Perturbation", *Journal of Guidance, Control, and Dynamics*, 2021. (Articles in Advance)

https://doi.org/10.2514/1.G005602